\documentclass[12pt,preprint]{aastex}
\usepackage{apjfonts}
\usepackage{epsfig}
\usepackage{amsmath}
\usepackage[colorlinks,linkcolor=blue,anchorcolor=green,citecolor=blue]{hyperref}

\begin{document}

\title{X-Ray Afterglow Plateaus of Long Gamma-Ray Bursts: Further Evidence for Millisecond
Magnetars}
\author{S. X. Yi$^{1,2}$, Z. G. Dai$^{1,2}$, X. F. Wu$^{3,4,5}$, and F. Y. Wang$^{1,2}$}

\affil{$^{1}$School of Astronomy and Space Science, Nanjing
University, Nanjing, China; dzg@nju.edu.cn}

\affil{$^{2}$Key laboratory of Modern Astronomy and Astrophysics
(Nanjing University), Ministry of Education, Nanjing 210093, China}

\affil{$^{3}$Purple Mountain Observatory, Chinese Academy of
Sciences, Nanjing 210008, China}

\affil{$^{4}$Chinese Center for Antarctic Astronomy, Chinese Academy
of Sciences, Nanjing, 210008, China}

\affil{$^{5}$Joint Center for Particle Nuclear Physics and Cosmology
of Purple Mountain Observatory-Nanjing University, Chinese Academy
of Sciences, Nanjing 210008, China}

\begin{abstract}
Many long-duration gamma-ray bursts (GRBs) were observed by {\it Swift}/XRT to have
plateaus in their X-ray afterglow light curves. This plateau phase
has been argued to be evidence for long-lasting activity of magnetar (ultra-strongly magnetized neutron stars)
central engines. However, the emission efficiency of such magnetars
in X-rays is still unknown. Here we collect 24 long GRB X-ray
afterglows showing plateaus followed by steep decays. We extend the
well-known relationship between the X-ray luminosity
${L_{\mathrm{X}}}$ and spin-down luminosity $L_{\mathrm{sd}}$ of
pulsars to magnetar central engines, and find that the initial rotation period
$P_{0}$ ranges from 1 ms to 10 ms and that the dipole magnetic field $B$ is centered around $10^{15}$ G. These
constraints not only favor the suggestion that
the central engines of some long GRBs are very likely to be rapidly
rotating magnetars but also indicate that the magnetar plateau emission efficiency in X-rays is close to 100\%.
\end{abstract}

\keywords{gamma-ray burst: general --- radiation mechanisms:
nonthermal --- stars: pulsars}

\section{Introduction}

Many gamma-ray bursts (GRBs) observed by \emph{Swift}/XRT
present plateaus prior to the subsequent power-law decay phase in their
early X-ray afterglows \citep{zhang06, Nousek06}. The
plateaus generally appear in 100-1000\,s since the GRB trigger with a typical slope $\alpha_1\sim0.5$ \citep{liang07}, where the flux of the plateau evolves as $F\propto t^{-\alpha_1}$. The distribution of the observed temporal decay slope $\alpha_2$ (defined in $F\propto t^{-\alpha_2}$) after the plateau, ranging from less than 1 to much greater than 1 (even to 10), is quite diverse. According to the standard model for GRB afterglows (for a recent complete reference see \cite{gao13b}), it is hard to understand some observed $\alpha_2$ with large values. Unless the plateau happens to be followed with the jet-like phase, which seems to be unlikely, the external shock models can not explain $\alpha_2>1.75$ (for details see next section). Therefore, it is now known that there are generally two types of plateaus. The first one is ``external plateau'', characterized by a normal decay ($\alpha_2\leq1.75$) after the plateau, which is currently understood as being due to energy injection into the external shock \citep{dai98a, dai98b, zhang01}. A tight correlation for X-ray plateaus between the break time $T_{b}$
and the corresponding X-ray luminosity $L_{\mathrm{X}}$ was recently
discovered by \cite{dainotti10} and \cite{xu12}, who selected the sample
with the slope of the follow-up decay phase generally less than 1.5. The second type is called ``internal plateau'', characterized by a steep decay ($\alpha_2>1.75$) after the plateau, which might be originated from internal dissipation of magnetic energy continuously blew out from the central engine \citep{troja07}. One possible candidate of
the central engine responsible for external energy injection as well as internally dissipative magnetic energy is an ultra highly magnetized and rapidly rotating
neutron star, which is also called magnetar \citep{thompson95}. The rotation energy of the magnetar can be tapped through magnetic dipole radiation (MDR) and/or relativistic leptonic wind \citep{dai04}. This speculation can be realized if the initial rotation period $P_{0}$ and dipole magnetic field $B$ of the
central neutron stars are found to be consistent with our expectation \citep{fan06, dai12}.

Since the magnetar model is almost the only successful model for internal plateaus\footnote{Matter-dominated energy injection is also possible for external plateaus, which is impulsively ejected and does not need long-lived central engine activity. It only requires a large variation in Lorentz factor. For large sample applications see \cite{Nousek06} and \cite{zhang06}.}, assuming dissipative magnetic energy is from MDR, one can derive the
initial period $P_{0}$ and magnetic field strength $B$ if the
spin-down luminosity $L_{\rm{sd}}$ and spin-down timescale
$T_{\rm{sd}}$ of the magnetar are known. \cite{rowlinson13} applied this method by assuming the emission
efficiency $\eta\equiv L_{\rm{rad}}/L_{\rm{sd}}=100\%$ to fit the observed X-ray plateaus, where
$L_{\rm{rad}}$ is the total bolometric luminosity in the $1-10^{4}$ keV in the cosmologically rest frame extrapolated from the observed X-ray luminosity $L_{\rm X}$ measured by {\it Swift}/XRT. \cite{zhang09} considered five remarkable
plateaus with sharp drops as a sample to discuss magnetars as the
central engine of GRBs, and found that the luminosity emitted in
X-ray band is a fraction of the total spin-down luminosity. In this paper, we
collect all {\it Swift} long GRBs with a steep decay after the plateau and
infer the stellar parameters based on the magnetar model. We assume
that the end time of the plateau phase corresponds to the spin-down time scale $T_{\rm{sd}}$
and that $\eta \lesssim 100\%$ is an adjustable parameter.

To more understand the physics behind $\eta$, we draw lessons from
persistent X-ray emission of normal pulsars. The dissipation of the rotation energy of a normal
pulsar to its persistent X-ray radiation could be similar to or the same as a millisecond
magnetar in a GRB, in which both spin down through magnetic dipole radiation.
Unlike GRB magnetars, the emission efficiency of a pulsar in X-ray can be directly calculated as the observed X-ray luminosity
$L_{\rm{X}}$ divided by the spin-down luminosity $L_{\rm{sd}}$.
In order to understand the mechanism by which the stellar rotation
energy is converted into X-ray emission, a tight correlation
between $L_{\rm{X}}$ and $L_{\rm{sd}}$ of pulsars has
been widely studied \citep{seward88, becker97, possenti02, cheng04}.
Because distinct components of X-ray emission have different origins for
normal pulsars, we here focus on the nonpulsed component of X-ray emission
from pulsar wind nebulae (PWNe). As stated above, since both millisecond magnetars in GRBs
and normal pulsars spin down through magnetic dipole radiation, we
assume that they have the same correlation between
$L_{\rm{X}}$ and $L_{\rm{sd}}$. Evidence for this assumption
is as follows. (1) \cite{gavriil08} found that the the dipolar
magnetic field of the young pulsar PSR J1846.0258 is about
$4.9\times 10^{13}$ G, which is higher than those of normal pulsars,
but lower than those of magnetars. Moreover, the detection of
magnetar-like emission from this pulsar suggests that there is a
continuum of magnetar-like activity throughout all neutron stars.
(2) \cite{vink09} found that the $L_{\mathrm{X}}$-$L_{\mathrm{sd}}$
correlation of the magnetar candidate anomalous X-ray pulsar 1E1547.0-5408
is similar to that of PWN pulsars. In this paper, therefore, we extend the
$L_{\mathrm{X}}$-$L_{\mathrm{sd}}$ correlation from normal pulsars
to magnetars, and obtain the spin-down luminosity of magnetars by
using the observed luminosity of a plateau.

The structure of the paper is as follows. In the next section, we
introduce the selection of the pulsar and plateau samples, and carry out
empirical fittings to the observed plateau light curves. The correlation in and between pulsars
and GRBs are calculated and discussed in section 3. Our conclusions and discussion
are presented in the last section.

\section{Sample Selection and Light Curve Fitting}
The nonthermal nonpulsed X-ray emission from rotation-powered
pulsars has been studied in the context of emission from PWNe. Here
we mainly investigate the nonpulsed X-ray emission from PWNe. We
exclude X-ray pulsars powered by accretion from binary companions,
and collect X-ray observational data of 101 pulsars with PWN from the
published literature \citep{possenti02, cheng04, li08, kargaltsev10}.
We find a correlation of $L_{\mathrm{X}}$-$L_{\mathrm{sd}}$ with the
101 PWN sample (see next section). This correlation also indicates the fraction of the
rotational energy loss of a pulsar going into X-ray emission.

X-ray plateaus are a common phenomenon in the afterglow
observations. Much work has been done for theoretical explanations
and statistic analysis for shallow decays \citep{dai98a,
dai98b, zhang01, dai04, liang07, dai12, yu07, yu10, dainotti10, xu12}.
In the external shock models, usually the decay slope after the plateau is $\alpha_2=(3p-2)/4$ if the environment is an interstellar medium (ISM) with a constant density, or sometimes (almost unlikely in X-ray after the plateau phase) $\alpha_2=(3p-1)/4$ if the environment is a stellar wind, where $p$ is the power-law index of the energy distribution of shock-accelerated electrons. The typical value of $p$ is about 2.3, however, it can range from 2.0 to 3.0 or even more smaller and larger. Therefore, the typical value of $\alpha_2$ is $\sim 1.2$ and the maximal allowable value by the model is 1.75. The possibility of the coincidence that the plateau happens to be followed by the jet-like phase is extremely small. Even in this case, $\alpha_2=\alpha_1+0.75\sim 1.3$ for an ISM environment and $\alpha_2=\alpha_1+0.5\sim 1.0$ for a wind environment, as long as the jet sideways expansion can be neglected. If the jet sideways expansion play the role, the value of $\alpha_2=p$ is typically $2.0 - 3.0$. As can be seen, the above values of $\alpha_2$ can not explain the large decay slope after the plateau observed in some GRBs. Internal plateaus with large $\alpha_2$ are thought to be due to magnetic energy dissipation at small radii, so that when the central engine ceases the decay timescale (equivalent to decay slope) is very short. In this
paper, we focus on internal plateaus and the criterion to be internal plateaus is $\alpha_2>1.75$. We have
collected 24 long duration GRB ($T_{90}\geq 2$ s) X-ray plateaus with this
judgement. Some of the GRBs in this sample have no
redshift measurements, and we adopt pseudo-redshift estimated by the $L_{\rm X}-T_b$ correlation for them \citep{dainotti10}. We suppose that the ending of an X-ray plateau corresponds to the spin-down time of the magnetar. The centrifugal force reduces as the magnetar spins down significantly, the magnetar collapses into a black hole due to the imbalance of the gravity and outward forces\footnote{Recently, \cite{zhang14} applied this scenario to interpret fast radio bursts (FRBs), a new type of cosmological
transients, although the physical nature is still
unknown.}. It is likely that the ending of the plateau, the spin-down and collapse of the magnetar coincidently happen at the same time.

we have collected 24 remarkable X-ray afterglow light curves in our sample. We apply a smooth broken
power-law and an extra power-law to fit the light curves. The fitting results are summarized
in Table 1. Generally speaking, the break (ending) time ($10^4-10^5$ s) of internal plateaus is
longer than that ($10^3-10^4$ s) of normal/external plateaus. The
break time of the internal plateau ($T_{b}$) is assumed to be the
spin-down time of a magnetar (${{T_{\rm{sd}}}}$), i.e.
\begin{equation}
{{T_{\mathrm{sd}}}}=\frac{{3{c^{3}}I}}{{{B^{2}}{R^{6}}\Omega _{0}^{2}}}=%
\frac{{3{c^{3}}I\,P_{0}^{2}}}{{4{\pi ^{2}}{B^{2}}{R^{6}}}},
\end{equation}%
where ${\Omega _{0}}=2\pi /{P_{0}}$ is the initial angular frequency,
$I$ is moment of inertia, $R$ is the stellar radius, and $c$ is the
speed of light. The isotropic X-ray luminosity at the break time $T_b$ is calculated by
\begin{equation}
L_{\mathrm{X}}=\frac{{4\pi D_{L}^{2}{F_{\rm X}}}}{{{{(1+z)}^{1-\beta
}}}},
\end{equation}%
where $z$ is the redshift, $D_{L}$ is the luminosity distance, $F_{\mathrm{X}%
}$ is the observed X-ray flux at the end time of the plateau phase, and
$\beta $ is the spectral index of the X-ray afterglow which can be
found from the \emph{Swift}/XRT website \citep{evans09}. The
spin-down luminosity of a pulsar/magnetar can be expressed as
\begin{equation}
L_{\mathrm{sd}}=\frac{{I\,\Omega _{0}^{2}}}{{2\,{T_{\mathrm{sd}}}}},
\end{equation}%
when $t\ll {{T_{\mathrm{sd}}}}$. With equations (1) and (3), we
obtain the initial period and the dipole magnetic field of the
pulsar/magnetar as
\begin{equation}
B=\left( \frac{{3{c^{3}}\,{I^{2}}}}{{2\,{R^{6}}L_{\mathrm{sd}}T_{\mathrm{sd}%
}^{2}}}\right) ^{1/2}
\end{equation}%
and
\begin{equation}
{P_{0}}=\left( \frac{{2{\pi ^{2}}\,I}}{L_{\mathrm{sd}}{\,{T_{\mathrm{sd}}}}}%
\right) ^{1/2}.
\end{equation}%
With the fitting results (see Table 1) and assuming $I=2\times 10^{45}$ g cm$%
^{2}$, $R=1\times 10^{6}$ cm, we can constrain the initial period ($P_{0}$) and the dipole
magnetic field strength ($B$) of the pulsar/magnetar.

\section{The $L_{\rm X} - L_{\rm sd}$  Correlation in Pulsars and GRBs}

It has been found that $L_{\mathrm{X}}$ and $L_{\mathrm{sd}}$ in pulsars have a
power-law relationship, but different authors have obtained different power-law
indices \citep{seward88, becker97, possenti02, cheng04, li08}. By
analyzing observed X-ray data of 101 PWN pulsars from the published
literature, we find that (see Figure 1)
$L_{\mathrm{X}}$ and $L_{\mathrm{sd}}$
have a tight correlation
\begin{equation}
L_{\mathrm{X}}=10^{-13.56\pm
1.90}\\\times \left(\frac{L_{\mathrm{sd}}}{{\rm erg}\,{\rm s}^{-1}}\right)^{1.28\pm
0.05}\,{\rm erg}\,{\rm s}^{-1}.
\end{equation}
Thus, the
corresponding conversion efficiency of the rotational energy of a
pulsar into nonpulsed X-ray emission is
\begin{equation}
\eta
=\frac{L_{\mathrm{X}}}{L_{\mathrm{sd}}}=10^{-13.56\pm1.90}\left(\frac{L_{\mathrm{sd}}}{{\rm erg}\,{\rm s}^{-1}}\right)^{0.28\pm
0.05},
\end{equation}
showing that the efficiency $\eta$ is
dependent on the spin-down luminosity.

The conversion efficiency of the rotational energy of a magnetar
into X-ray emission, in order to interpret X-ray plateaus, is unknown.
Some papers, such as \cite{rowlinson13}, generally adopted the
efficiency of the rotational energy into the $1 - 10^4$ keV emission as $100\%$ in their calculations.
However, their extrapolation from X-ray to $1 - 10^4$ keV is based on the X-ray spectral index, which may not be applicable beyond the XRT band.
In this paper we consider X-ray plateaus followed by steep decays as central engine
afterglows from magnetars, and assume that such magnetars and
rotation-powered pulsars have the same mechanism that X-ray emission are from internal dissipation of Poynting flux. Evidence for this
assumption comes from the possible fact that pulsars and magnetars
may have the same $L_{\mathrm{X}}$-$L_{\mathrm{sd}}$ correlation
\citep{gavriil08, vink09}, that is, the $L_{\mathrm{X}}$ and $L_{\mathrm{sd}}$ of the magnetar candidate, the anomalous X-ray pulsar 1E1547.0-5408,
satisfy the correlation in PWN pulsars. Therefore, we extend the correlation
of $L_{\mathrm{X}}$ and $L_{\mathrm{sd}}$ from rotation-powered pulsars
to magnetars. The corresponding conversion efficiency of
the rotational energy of a magnetar into X-ray emission is also
given by equation (7).

The spin-down
luminosity during X-ray plateaus can be calculated by
equation (6), where
$L_{\mathrm{X}}$ is the luminosity at the end of the plateau (see
Table 1). There are some GRBs without redshift measurement. In these cases, their
redshifts can be estimated by the correlation between $L_{\rm X}$
and $T_{b}$ from \cite{dainotti10}. With the derived spin-down
luminosity, we can further calculate the initial period $P_{0}$ and the dipole
magnetic field strength $B$ of a magnetar with equations (4) and
(5). Table 1 shows that the derived initial spin period of the
magnetars ranges from 1 to 10 ms, which is well consistent with
the values expected in the magnetar formation hypothesis. The dipole
magnetic field $B$ of Table 1 is in the range of $10^{14}-10^{15}$ G,
which is consistent with the magnetic field of soft gamma-ray repeaters and anomalous X-ray pulsars.

Figure 2 shows the magnetic field and spin period for both long
and short GRB candidates. The magnetars could be roughly divided
by $B=5\times10^{15}$ G into two different samples, short GRB
candidates above the line and long GRB candidates below the line.
One caveat is that there are some GRBs with extended emission included
in the sample plotting Figure 2. Because their distribution is similar to
that of the short ones \citep{rowlinson13,gompertz13}, we consider
them as one subset of the short GRB candidates. Compared with the long GRBs,
the short GRB candidates tend to have higher magnetic fields. From our statistics, we find the initial spin period
mainly in the range $1 - 10$ ms and the dipole magnetic field in the
range $5\times10^{14}-5\times10^{15}$ G for the long GRB
magnetars. These values are all reasonable, implying that
internal plateaus could be powered by a central spinning-down magnetar.

\section{Conclusions and Discussion}

The X-ray plateaus can be explained as being due to continuous
energy injection from central engines after the prompt bursts and
rapidly rotating, ultra-strongly magnetized pulsars are good
candidates of such GRB central engines. In this paper, we have
collected 24 remarkable long-GRB X-ray plateaus followed by sharp
drops. We assumed that the X-ray plateaus are powered by internal
magnetic energy dissipation of Poynting flux from a magnetar and
the sudden drop is caused by the spin-down and collapse of the magnetar. On the other hand, we gathered the X-ray observational data on $L_{\rm X}$ and $L_{\rm sd}$ of 101 PWN
pulsars from the published literature, and fitted them
with a power law function (Figure 1), $L_{\mathrm{X}}={10^{(-13.56\pm
1.90)}}(L_{\mathrm{sd}}/{\rm erg}\,{\rm s}^{-1})^{(1.28\pm
0.05)}\,{\rm erg}\,{\rm s}^{-1}$. We assumed
that magnetars and rotation-powered pulsars may experience a common internal
dissipation mechanism. Thus, we extended the correlation of $L_{
\mathrm{X}}$ and $L_{\mathrm{sd}}$ from rotation-powered pulsars to
X-ray plateaus. We find that for the magnetar candidates in the 24 long GRBs, the initial period $P_{0}$
is about 1 to 10 ms, while the dipole magnetic field strength $B$ is
in the range of $10^{14}$ to $10^{15}$ G. This result implies that the central
engines of some long GRBs are millisecond magnetars.

Millisecond magnetars are not only proposed as the central engines
of some long GRBs, but also they may survive from some binary neutron
star mergers that power short GRBs. The long-lasting activity of the
central magnetars have been suggested to interpret the X-ray flares
and plateaus following some short GRBs %
\citep{dai06,fan06,rowlinson10, rowlinson13} and
the statistical properties of X-ray flares from both long and short GRBs %
\citep{wangfy13}. Recently, such a survived massive millisecond magnetar scenario has been studied
to predict a bright multi-wavelength afterglow \citep{gao13a}
and invoked to explain an unusual energetic transient PTF11agg \citep{wang13,wu14}.
We therefore suggest that millisecond magnetars play an important role in both
long and short GRBs.

%*******************************************************************************************************************
%*******************************************************************************************************************
\section*{Acknowledgements}
We acknowledge the use of the public data from the {\it Swift} archives and the data supplied by the UK {\it Swift} Science Data Centre at the University of Leicester.  We
thank Bing Zhang, Xiang-Yu Wang, Yong-Feng Huang, Wei Wang, Xuan Ding and Ling-Jun Wang
for useful comments and helps. This work is supported by the National
Basic Research Program of China (973 Program, grants 2014CB845800 and 2013CB834900) and
the National Natural Science Foundation of China (grants 11373022, 11322328,
11103007, and 11033002). XFW acknowledges support by the One-Hundred-Talents Program and the Youth Innovation Promotion Association of Chinese Academy of Sciences.

%*******************************************************************************************************************
%*******************************************************************************************************************

\clearpage
\begin{figure}[t]
\centering
\includegraphics[angle=0,scale=0.5,width=0.65\textwidth,height=0.40
\textheight]{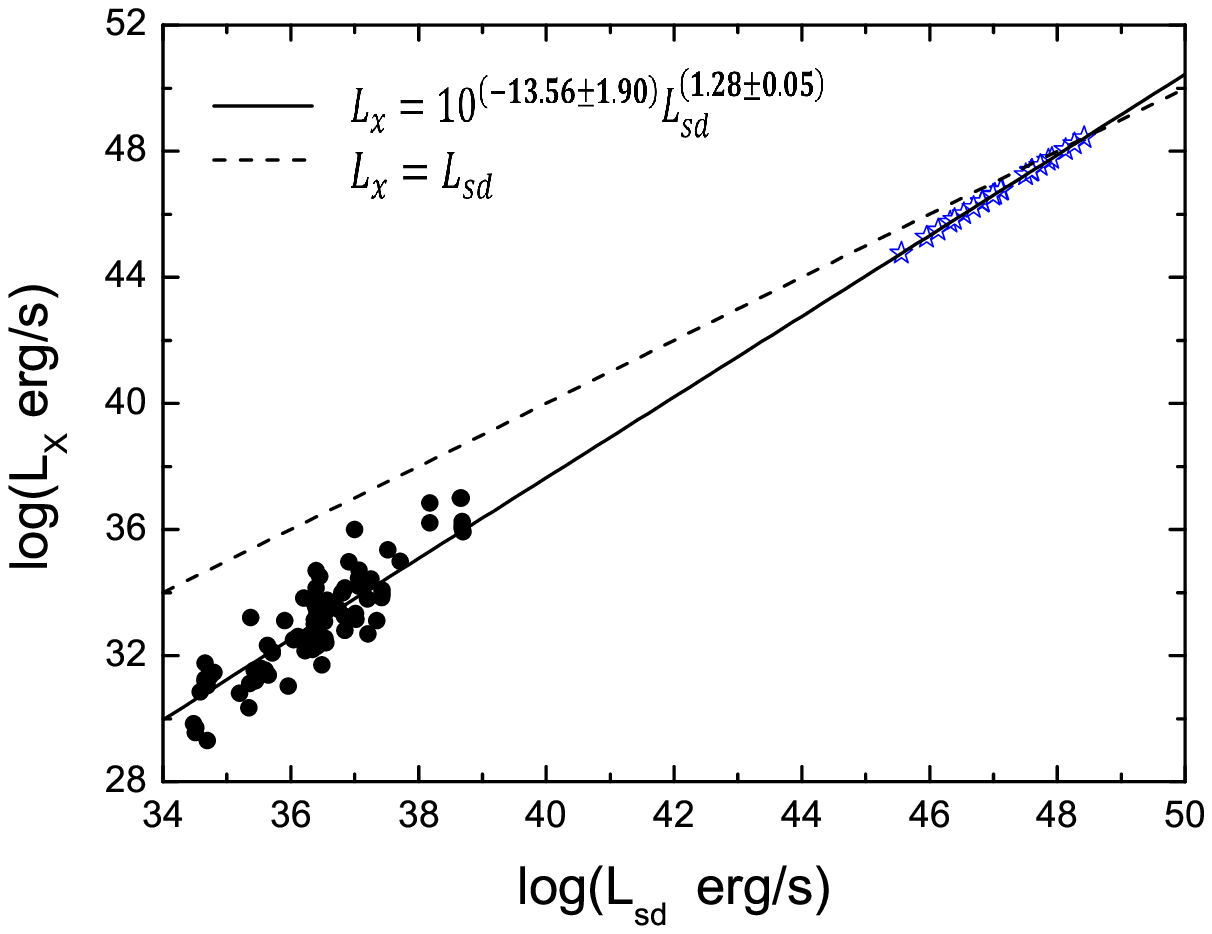}\caption{The $L_{\rm X} - L_{\rm sd}$ correlation in pulsars (solid black dots) and
long GRBs (open blue stars). The solid line corresponds to the best fit for pulsars, while the dashed line is $L_{\rm sd}=L_{\rm X}$. Luminosities are in units of erg s$^{-1}$.}
\end{figure}

\clearpage
\begin{figure}[t]
\centering
\includegraphics[angle=0,scale=0.5,width=0.70\textwidth,height=0.40
\textheight]{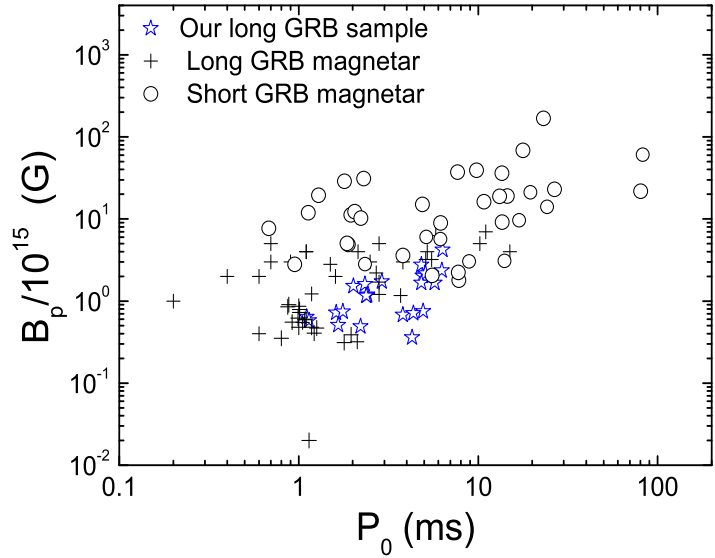} \caption{Magnetic field and spin period of the magnetar candidates in GRBs. Black crosses
are the magnetar candidates in long GRBs taken from \cite{yu10}, \cite{lyons10}, \cite{dall'Osso11}, \cite{bernardini12}.
Open circles are the magnetar candidates in short GRBs identified by \cite{rowlinson13}, \cite{gompertz13}. Open blue stars
stand for our sample. Magnetars in short GRBs tend to have higher magnetic fields.}
\end{figure}

%*******************************************************************************************************************
%*******************************************************************************************************************
\clearpage
\begin{deluxetable}{ccccccccccccccccccccccccc}
\tabletypesize{\scriptsize}
\tablecaption{Fitting results of X-ray plateaus and derived parameter values for magnetar candidates }
\tablewidth{0pt}
%\tabletypesize{\tiny}

\tablehead{ \colhead{GRB}
&\colhead{$z$}
&\colhead{$\alpha_{1}$}
&\colhead{$\alpha_{2}$}
&\colhead{$T_{b}$\tablenotemark{a}}
&\colhead{$F_{x}$\tablenotemark{b}}
&\colhead{$L_{x}$\tablenotemark{c}}
&\colhead{$L_{sd}$\tablenotemark{c}}
&\colhead{$P_{0}$\tablenotemark{d}}
&\colhead{$B$\tablenotemark{e}}
&\colhead{Refs.}}

\startdata
060413  &   0.61$^{*}$  &   2.2E-4  $\pm$   0.04    &   3.09    $\pm$   0.09    &   24224.7 $\pm$   456.7   &   22.37   $\pm$   0.65    &   2.35     $\pm$  0.07    &   6.62    &   4.96    &   2.04    &   ... &\\
060605  &   3.8 &   0.42    $\pm$   0.05    &   1.89    $\pm$   0.05    &   7254.9  $\pm$   252.6   &   8.07    $\pm$   0.31    &   110.17  $\pm$    4.30   &   133.85  &   2.02    &   1.52    &   1   &\\
060607A &   3.082   &   0.31    $\pm$   0.03    &   3.60    $\pm$   0.07    &   12258.8 $\pm$   190.4   &   56.24   $\pm$   1.76    &   260.60  $\pm$    8.14   &   262.26  &   1.11    &   0.64    &   2   &\\
060923C &   1$^{*}$ &   0.46    $\pm$   0.06    &   1.79    $\pm$   0.23    &   179436.3    $\pm$   19430.2 &   0.24    $\pm$   0.03    &   0.18     $\pm$  0.03    &   0.90    &   4.94    &   0.75    &   ... &\\
070110  &   2.352   &   0.18    $\pm$   0.05    &   9.79    $\pm$   0.55    &   20714.4 $\pm$   218.7   &   10.94   $\pm$   0.43    &   50.65   $\pm$    2.00   &   72.94   &   1.62    &   0.72    &   3   &\\
070429A &   1.3$^{*}$   &   0.34    $\pm$   0.04    &   8.87    $\pm$   4.73    &   592515.7    $\pm$   67316.5 &   0.07    $\pm$   0.01    &   0.06     $\pm$  0.01    &   0.36    &   4.28    &   0.36    &   ... &\\
070611  &   2.04    &   -0.77   $\pm$   0.52    &   1.78    $\pm$   0.18    &   29274.8 $\pm$   4800.6  &   0.43    $\pm$   0.07    &   1.01    $\pm$    0.16   &   3.42    &   6.27    &   2.35    &   4   &\\
070721B &   3.626   &   0.65    $\pm$   0.04    &   2.32    $\pm$   0.10    &   8819.7  $\pm$   244.1   &   10.18   $\pm$   0.38    &   58.64   $\pm$    2.17   &   81.78   &   2.34    &   1.60    &   5   &\\
071118  &   1.24$^{*}$  &   0.39    $\pm$   0.08    &   2.51    $\pm$   0.29    &   12500.3 $\pm$   478.1   &   9.27    $\pm$   0.44    &   5.90     $\pm$  0.28    &   13.60   &   4.82    &   2.76    &   ... &\\
080703  &   1.5$^{*}$   &   0.58    $\pm$   0.02    &   2.60    $\pm$   0.20    &   24295.5 $\pm$   1989.6  &   2.72    $\pm$   0.55    &   2.41     $\pm$  0.49    &   6.76    &   4.90    &   2.02    &   ... &\\
081029  &   3.848   &   0.42    $\pm$   0.06    &   2.44    $\pm$   0.11    &   16791.8 $\pm$   1219.5  &   5.21    $\pm$   0.10    &   23.77   $\pm$    2.48   &   40.39   &   2.41    &   1.19    &   6   &\\
090205  &   4.7 &   0.52    $\pm$   0.11    &   2.11    $\pm$   0.19    &   17493.8 $\pm$   1251.4  &   0.98    $\pm$   0.08    &   22.60   $\pm$    1.96   &   38.83   &   2.41    &   1.17    &   7   &\\
090308  &   2.38$^{*}$  &   0.34    $\pm$   0.36    &   2.94    $\pm$   0.21    &   128650.9    $\pm$   100000.0    &   0.06    $\pm$   0.02    &    0.54   $\pm$   0.21    &   2.11    &   3.81    &   0.68    &   ... &\\
090807  &   1.44$^{*}$  &   -0.08   $\pm$   0.07    &   1.79    $\pm$   0.10    &   9368.0  $\pm$   669.0   &   2.47    $\pm$   0.17    &   4.22     $\pm$  0.29    &   10.46   &   6.34    &   4.20    &   ... &\\
100219A &   4.5 &   0.18    $\pm$   0.15    &   2.17    $\pm$   0.33    &   23527.5 $\pm$   2365.4  &   3.45    $\pm$   0.37    &   34.48   $\pm$    3.68   &   54.01   &   1.76    &   0.74    &   8   &\\
100508A &   1.24$^{*}$  &   0.29    $\pm$   0.07    &   2.61    $\pm$   0.12    &   22563.7 $\pm$   1055.9  &   4.43    $\pm$   0.27    &   2.42     $\pm$  0.15    &   6.77    &   5.08    &   2.17    &   ... &\\
100614A &   1.21$^{*}$  &   0.28    $\pm$   0.06    &   2.11    $\pm$   0.22    &   153270.0    $\pm$   12469.8 &   0.40    $\pm$   0.03    &   0.31     $\pm$  0.02    &   1.36    &   4.36    &   0.71    &   ... &\\
100906A &   1.727   &   0.70    $\pm$   0.02    &   2.07    $\pm$   0.04    &   11697.4 $\pm$   260.2   &   11.52   $\pm$   0.35    &   23.26   $\pm$    0.71   &   39.71   &   2.91    &   1.73    &   9  &\\
111022B &   2.5$^{*}$   &   -0.04   $\pm$   0.15    &   2.65    $\pm$   0.70    &   48625.1 $\pm$   8148.5  &   0.23    $\pm$   0.04    &   0.67     $\pm$  0.11    &   2.47    &   5.73    &   1.66    &   ... &\\
111209A &   0.677   &   0.58    $\pm$   0.00    &   5.47    $\pm$   0.04    &   16116.0 $\pm$   33.4    &   958.78  $\pm$   5.19    &   169.02  $\pm$    0.91   &   186.99  &   1.14    &   0.58    &   10  &\\
120320A &   3.14$^{*}$  &   0.02    $\pm$   0.09    &   4.25    $\pm$   0.60    &   82527.1 $\pm$   8089.4  &   0.39    $\pm$   0.06    &   3.89     $\pm$  0.58    &   9.82    &   2.21    &   0.49    &   ... &\\
120326A &   1.798   &   -0.38   $\pm$   0.03    &   1.86    $\pm$   0.05    &   44331.0 $\pm$   1254.1  &   9.47    $\pm$   0.21    &   17.74   $\pm$    0.40   &   32.14   &   1.66    &   0.51    &   11  &\\
120521C &   6   &   0.21    $\pm$   0.11    &   2.12    $\pm$   0.26    &   17204.3 $\pm$   2589.3  &   0.84    $\pm$   0.11    &   23.74   $\pm$    3.18   &   40.35   &   2.38    &   1.16    &   12  &\\
130315A &   2.04$^{*}$  &   0.07    $\pm$   0.12    &   2.41    $\pm$   0.30    &   34498.6 $\pm$   3733.4  &   0.60    $\pm$   0.07    &   1.60     $\pm$  0.19    &   4.91    &   4.83    &   1.66    &   ... &\\
\enddata
\tablenotetext{*}{The redshifts of those GRBs are constrained by the correlation of $L_{x}$ and $T_{b}$ from \cite{dainotti10}.}
\tablenotetext{a}{In units of second.}
\tablenotetext{b}{In units of $10^{-12}$ erg cm$^{-2}$s$^{-1}$.}
\tablenotetext{c}{In units of $10^{46}$ erg/s.}
\tablenotetext{d}{In units of ms.}
\tablenotetext{e}{In units of $10^{15}$ G.}
\tablerefs{
(1)  \cite{ferrero09};
(2)  \cite{molinari07};
(3)  \cite{jaunsen07};
(4)  \cite{thoene07};
(5)  \cite{malesani07};
(6)  \cite{cucchiara08};
(7)  \cite{fugazza09};
(8)  \cite{groot10};
%(9)  Campana et al. 2010;
(9) \cite{tanvir10};
(10) \cite{vreeswijk11};
(11) \cite{tello12};
(12) \cite{tanvir12};
}
\end{deluxetable}

%*******************************************************************************************************************
%*******************************************************************************************************************

\end{document}